# Method for Flow Measurement in Microfluidic Channels Based on Electrical Impedance Spectroscopy

Nima Arjmandi [†, a, b], Chengxun Liu [a], Willem Van Roy [a], Liesbet Lagae [a, b], Gustaaf Borghs [a, b]

[a] *IMEC, Kapeldreef 75, 3001 Leuven, Belgium*

[b] *Department of Physics and Astronomy, Katholieke Universiteit Leuven, Celestijnenlaan 200 D, 3001 Leuven, Belgium*

[†] Telephone: +32 16 28 77 66

Fax: +32 16 28 10 97

Electronic mail: Arjmandi@imec.be

We have developed and characterized two novel micro flow sensors based on measuring the electrical impedance of the interface between the flowing liquid and metallic electrodes embedded on the channel walls. These flow sensors are very simple to fabricate and use, are extremely compact and can easily be integrated into most microfluidic systems. One of these devices is a micropore with two tantalum/platinum electrodes on its edges; the other is a micro channel with two tantalum /platinum electrodes placed perpendicular to the channel on its walls. In both sensors the flow rate is measured via the electrical impedance between the two metallic electrodes, which is the impedance of two metal-liquid junctions in series. The dependency of the metal-liquid junction impedance on the flow rate of the liquid has been studied. The effects of different parameters on the sensor's outputs and its noise behavior are investigated. Design guidelines are extracted and applied to achieve highly sensitive micro flow sensors with low noise.

# 1. Introduction

Flow sensors together with pumps, mixers, valves, channels and filters are the key components of microfluidic systems. The most available solutions for sensing the flow rate of fluids in microfluidics are commercially available decimeter sized off-chip micro flow sensors. However, these have a large dead volume, limited sensitivity, large size and difficulties in interfacing with microfluidic devices. A small integratable micro flow sensor is essential, especially considering the size reduction trend, and also in lab on chip and micro total analysis systems (µ-TAS).

To address this need, researchers have investigated different phenomena to sense the flow rate of fluids on the micrometer scale and fabricated different devices. A major group of these flow sensors are based on heat transfer detection (Chen et al. 2003; Ernst et al. 2002; Glanninger et al. 2000; Meng et al. 2001;Okulan et al. 1998; Sabate et al. 2004; Wu et al. 2001). Although these sensors have high sensitivity, they need a complicated massive structure difficult to integrate in a microfluidic chip. Mechanical flow sensors that use drag force (Attia et al. 2009; Fan et al. 2002) and differential pressure flow sensors (Alvarado et al. 2009), also tend to have bulky and complicated structures and are usually not sufficiently sensitive. Other groups of micro flow sensors include time of flight electrochemical sensors (Wu and Ye 2005; Wu et al. 2001), optical sensors (Lien and Vollmer 2007; Nguyen et al. 2005), atomic emission flow sensors (Nakagama et al. 2003), streaming potential (Caldwell et al. 1986; Kim et al. 2006) and ion sensitive field effect transistor flow sensors (Truman et al. 2006). Unfortunately, most of these also lack high sensitivity and are bulky with complicated structures, complicating their integration into microfluidic devices. In addition they may suffer from a long response time, and high hysteresis.

In this paper we are introducing another phenomenon for micro flow sensing; namely, the dependence of the electrical impedance of a metal-liquid interface on the flow rate of the liquid. We have fabricated two different kinds of flow sensors based on this phenomenon, which can be very small, highly sensitive, simply structured and integratable. Moreover, the dependency of the output of these novel sensors on different parameters, the sensitivity, and noise behavior have all been investigated; different questions about this phenomenon have been answered and design guide lines have been developed.

# 2. Working principles

At the interface between a liquid and a solid, accumulation or depletion of ionic or electronic charge almost always exists. This accumulation and depletion gives rise to an electric double layer (EDL). The EDL consists of a layer of immobilized charges on the solid's surface which is called Stern layer and an accumulation or depletion of mobile charges in the liquid which is called diffuse layer (Li 2004). If we put two metal electrodes on the walls of a microfluidic channel in contact with the liquid and orthogonal to the flow direction (fig. 3), the equivalent electrical circuit of these electrodes and the liquid (fig. 1) will contain: two capacitors ($C_s$) representing the Stern layer, two capacitors ($C_d$) representing the diffuse layers, a resistor ($R$) representing the resistivity of the bulk liquid, a capacitor ($C$) originating from the two metal electrodes as parallel plates and the liquid and the substrate as dielectrics, two other resistors ($r$) representing the electron transfer between the electrodes and the liquid, and two Warburg impedances ($W$) representing the diffusion limitation of mass transfer (Hong et al. 2004; Orazem and Tribollet 2008). Considering the different compositions of the EDL near the metallic electrodes and the EDL near the channel walls, flow of liquid in the channel will change the values of $r$, $W$ and $C_d$ in the following ways and we can measure this change by measuring the total impedance between the two electrodes:

1. The flow will drag some of the ions of the metal's diffuse layer (usually negative ions) downstream and replace them by the ions coming from upstream (usually positive ions of the diffuse layer of the silicon walls). So, the charge of the diffuse layer of the electrode will change and $C_d$ will change consequently (Ayliffe and Rabbitt 2003).

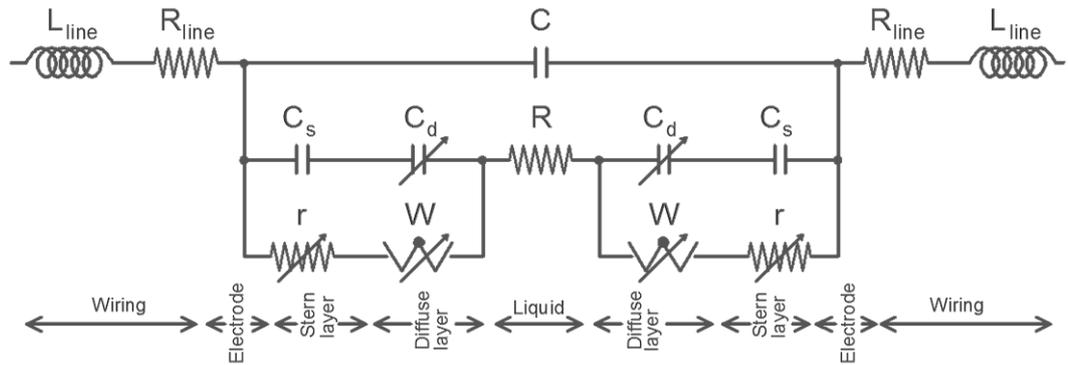

**Figure 1.** Schematic diagram of the equivalent circuit of the sensor. The components which are changing by the flow rate are shown as variable components ($r$, $W$ and $C_d$). $R_{Line}$ and $L_{Line}$ are the parasitic effects of the measurement setup. Bellow the circuit, the origin of each segment of the circuit is mentioned.

2. If the applied voltage to the metal-liquid junction is sufficiently high, current will be limited by the amount of reacting ions supplied to the surface by diffusion or convection or flow (mass transfer limit). When the current is well below mass transfer limit, it is controlled by the voltage; and flow of the liquid will not change the Warburg impedance ($W$). But, when the current is mass transfer limited, using drift-diffusion current equation (Morrison 1980) and the equation of convective-diffusive mass transport (Levich 1962) one can show that the DC impedance of the interface ($W + r$) is decreasing with the flow rate (Collins and Lee 2003).

## 3. Experiment

We have fabricated two different types of devices; the first device is a micrometer sized pore in a silicon membrane with two electrodes at its edges (fig 2a). These electrodes are electrically isolated from each other and from the liquid, except at the edges of the pore where they are in contact with liquid. There are two millimeter sized chambers on the two sides of the membrane (fig. 2). These chambers are filled with the liquid which flows through the pore from one chamber to the other.

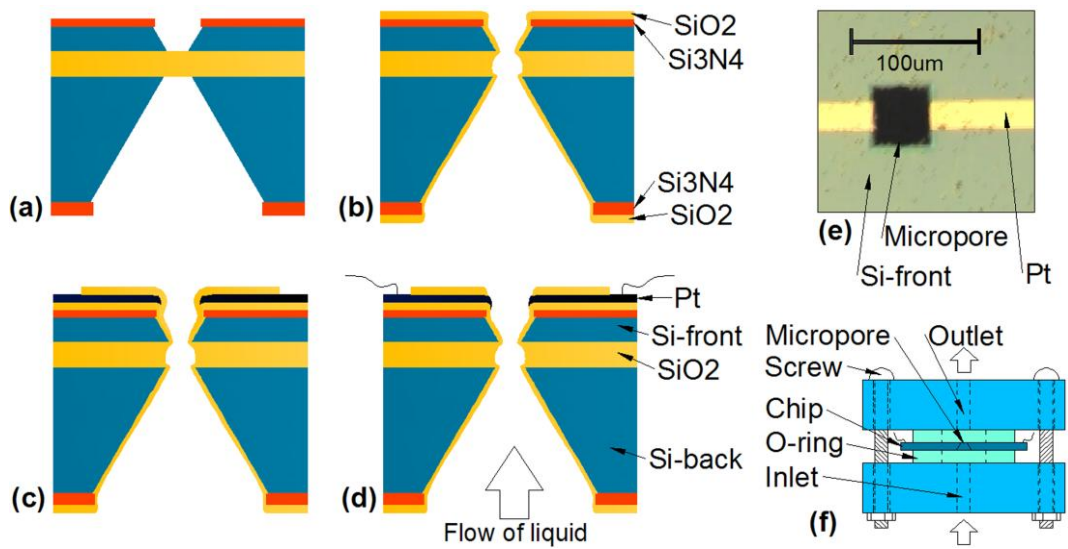

**Figure 2.** a,b,c and d) Schematic cross section of the micropore device fabrication process (not to scale). e) Top view of a 50x50 μm$^2$ micropore and the metal electrodes (the scale bar on the microscope picture is 100 μm). f) Schematic of the packaged micropore device.

We have fabricated our micropore device on silicon-on-insulator wafer consisting of a 720 nm Si (001) top layer and a 1 μm thick buried oxide (BOX) on a 700 μm thick Si (001) substrate. After cleaning the wafer, low pressure chemical vapor deposition (LPCVD) has been used to deposit 150 nm of silicon nitride on both sides of the wafer. Then, a window is defined in the front side's nitride by e-beam lithography and reactive ion etching (RIE). Anisotropic etching using a 33% KOH solution at 40°C, (approx. 3 min) is used to etch a pit on the front side with bottom dimension of about 50 μm. In the next step, a 780x780 μm$^2$ window is opened in the backside nitride by RIE and a 10 % KOH solution used to etch a pit all the way through the 700 μm thick back side's silicon in 12 hours at 80°C (fig. 2.a), using the BOX layer as an etch stop. Then the BOX layer is removed by immersing in buffered HF (BHF) for 10 min. In order to prevent any unwanted electrical conduction through the substrate, a 1.2 μm thick layer of silicon dioxide is sputtered on the front side and back side of the device (fig. 2.b).

The metallic electrodes are patterned on this oxide layer by lift-off. As spray coating of photoresist (Yu et al. 2006) was not available, we have developed the following spin coating process to make a homogeneous layer of the resist around the micropore. Samples were heated at 90° C on hot plate for 15 sec. Then, AZ5214 negative resist (MicroChemicals) was purged on them while they were on the hot plate and remained on the hot plate for 5 sec. Then the resist was spun for 30 sec at 4000 rpm and baked on a hot plate for 60 sec at 90°C. The 20 nm Ta/200 nm Pt electrodes were then defined by classical UV lithography and lift-off. After fabrication of the electrodes, 1 μm thick silicon dioxide was sputtered on them, while the contact pads were shadow masked (fig. 2.c). Finally, the tips of the electrodes (at the edge of the pore) were exposed again by RIE through the backside of the pores, with the wafer mounted face-down in the etching chamber (fig. 2.d). In back end, 200 μm thick lead coated cupper wires were soldered to the contact pads at 410°C and the devices were packaged between 2 pieces of Plexiglas with inlets and outlets drilled in them. These Plexiglas are sealed to the silicon die by PDMS O-rings and pressed to the chip by four screws (fig. 2.f). These O-rings and Plexiglas form two liquid filled chambers on the two sides of the pore.

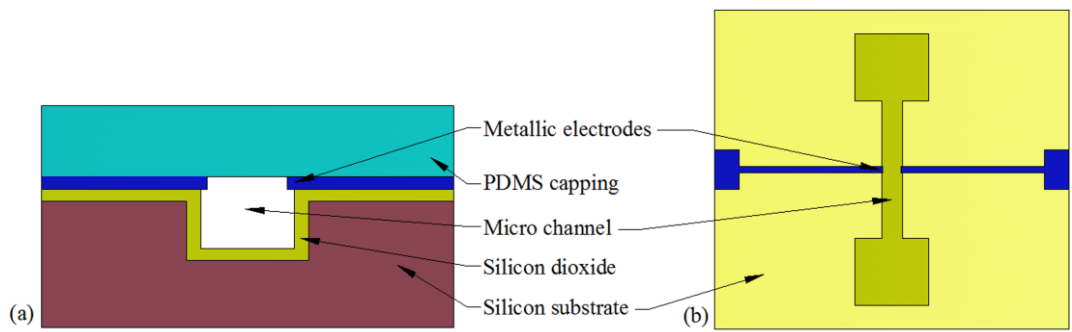

**Figure 3.** Schematic of the micro channel device (not to scale). a); cross section of the device along the electrodes. b) Top view of the device.

The second type of flow sensor devices that we have made consists of a micrometer sized channel in silicon with two metallic electrodes perpendicular to the channel and perpendicular to the direction of flow as well (fig. 3). These electrodes are electrically isolated from each other and they are in contact with the liquid just at their ends. After cleaning the Si substrate, 1 μm of silicon dioxide was sputtered on it. Then we used UV lithography and lift-off to define a Ta/Pt (20/200 nm) hard mask which is used for etching 30 μm deep channels in the Si substrate by RIE. A layer of Ta/Cu (20/200 nm) with electrode's pattern was defined by lift-off using the same special lithography process mentioned above. Then, this layer has been used as a mask to etch the Ta/Pt layer by ion milling.

In back end, lead coated cupper wires were soldered to the contact pads. Then, the silicon chip together with a sheet of PDMS was treated with $O_2$ plasma of for 15 min at 100 W, the sample's surface wetted by a water droplet and the PDMS put on the die to cap the channels. PDMS bonding was finished by baking the assembly for 3 min at 70°C, followed by 2 min at 90°C, 2 min at 100°C, 2 min at 110°C, and 7 min cool-down. Finally, inlet and outlet plastic tubes were glued to holes that were punched in the PDMS on the inlet and outlets.

A range of different solutions with different compositions and concentrations is injected in to the devices by a calibrated syringe pump (Harvard Apparatus). The impedance between the electrodes is measured by PGSTAT302N (Metrohm Autolab) at different frequencies and voltages with 1 sec integration time and 1 sec stabilization time before each measurement. Impedance analysis, simulations and curve fittings are done using ZView (Scriber).

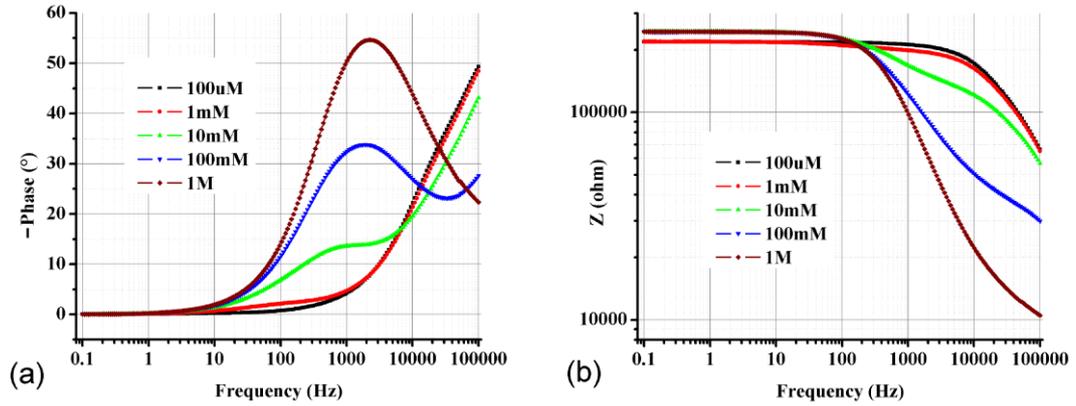

**Figure 4.** a) Phase plot and b) Bode plot of impedance obtained by a micropore device made of a 80x80 μm$^2$ hole and 40 μm wide electrodes filled with different concentrations of calcium chloride in water. The applied voltage was a 50 mV RMS sinusoidal wave with frequencies between 0.01 Hz and 100 kHz.

## 4. Results

We have measured the electrical impedance between the two electrodes in different frequencies, voltages, flow rates, solutions and concentrations. Figure 4 shows the impedance between the electrodes of a micropore device in different $CaCl_2$:$H_2O$ solutions. According to the measurement results shown in figure 4 and in agreement with the device geometry, at very low frequencies, the leakage current is dominant (r and W in Fig. 1) and the concentration and frequency dependencies are very small. At higher frequencies, EDL capacitances ($C_s$ and $C_d$) are introducing a phase shift and the impedance starts to decrease. This region is where the phase curve (fig. 4.b) starts to rise. By further increasing the frequency, the liquid's resistivity (R) limits the impedance and the phase curve decreases again. At higher frequencies, the cell capacitance (C) dominates and phase shift increases again and the impedance reduces further.

By introducing a pressure driven flow, imaginary and real parts of the impedance are decreasing (fig. 5). At low flow rates, this decrease is almost linear. But, at higher flow rates, gradually the sensitivity of sensor decreases and the impedance saturates. The dynamic range of the sensor depends on the measurement frequency, the solution and the geometry of the device; by increasing the measurement frequency or increasing the channel width, the linear region extends, but the sensitivity decreases.

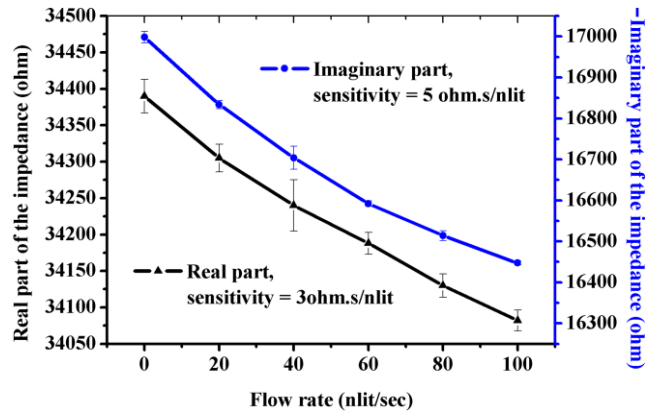

**Figure 5.** Imaginary (■) and real parts (▲) of the impedance as a function of flow rate 165 mM $CaCl_2:H_2O$ solution in a micropore device made of an 80x80 $\mu m^2$ pore and 40 μm wide electrodes. Measurements were done at 5 kHz with 50 mV RMS sinusoidal voltage.

This sensor is successfully used to measure the flow rate of water, calcium chloride solutions, potassium chloride solutions and glucose solutions. This flow sensor can work with different ionic concentrations ranging from nonionic solutions to pure water, or a few molar ionic solutions. Like most of the other flow sensors, the sensor's parameters are depending on the liquid. Fortunately, in the case of having unknown liquids, it's possible to extract some information on the liquid's composition by electrochemical spectroscopy (Delaney et al. 2007) by the sensor itself. But, considering the approximately linear output of the sensor in its working range, we can calibrate the sensor by recording its output at different flow rates; the normalized output has no significant dependence on the ionic concentration of the solution (fig. 6.a).

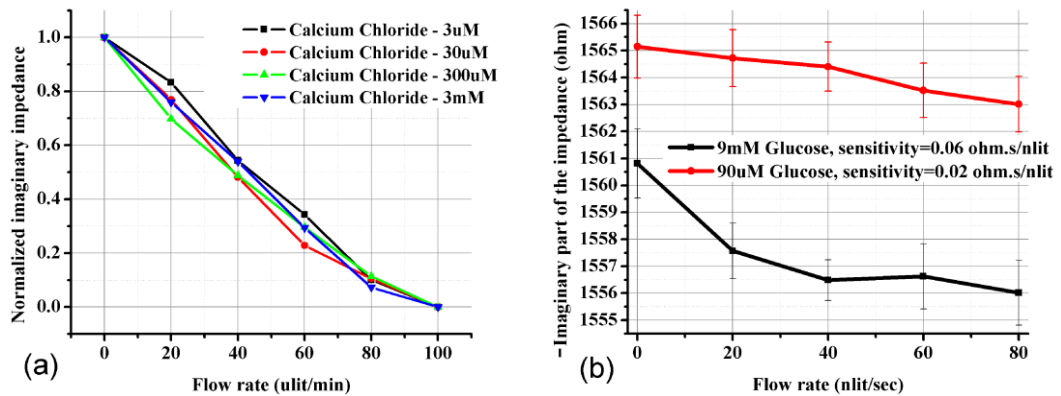

**Figure 6.** a) Normalized sensor response in different ionic concentrations or $CaCl_2:H_2O$ ranging from 165 μM to 165 mM. b) Sensor response to nonionic solutions (5 mM and 500 mM glucose:water). Results obtained in micro channel device with a 50 μm wide channel and 50 μm wide electrodes. Applied voltage was 5 kHz sinusoidal wave with 50 mV RMS.

In nonionic solutions, the sensitivity of the sensor is quite low (fig. 6.b) and the noise level is quite high. In the most of the designs and conditions, highest sensitivity and signal to noise ratio was obtained with low concentration ionic solutions.

To analyze effects of the different frequencies on the sensitivity, we have defined the sensitivity (Ω s / L) of the devices as:

$$sensitivity = \frac{\text{Im}\{Z(Q_1)\} - \text{Im}\{Z(Q_2)\}}{Q_1 - Q_2}$$

where $Q_1$ and $Q_2$ are two different flow rates, $Z$ is the measured impedance between the electrodes. The sensitivity of different devices using different solutions was measured as a function of frequency. In all of the measurements a peak in the sensitivity vs. frequency curve is observed (fig 7.b). An explanation is that at very low frequencies, the total impedance is dominated by behavior of the ions far away from the surface (Wu and Ye 2005) ($R$ in the equivalent circuit, figures 1 and 4) where flow of liquid does not introduce much change in the ionic distribution. On the other hand at high frequencies, the total impedance is sensitive to changes in the ions in close proximity to the surface ($C_s$ in the equivalent circuit, figures 1 and 4) (Wu and Ye 2005). These ions are mostly in the Stern layer and their distribution is not changing with flow. But, at moderate frequencies; the total impedance is mostly determined by the ions in the diffuse layer, and $C_d$ and $W$ are dominant (figures 1 and 4). Because the flow introduces the biggest change in the diffuse layer, the sensitivity of the sensor has a maximum at these frequencies.

The concentration which results in the highest sensitivity depends on the frequency as well; higher frequencies are giving the highest sensitivity at higher concentrations (fig. 7.a). We believe that at high concentrations the diffuse layer which senses the flow of liquid is more confined to the surface and we need higher frequencies to be sensitive to the close proximity of the surface. Also, the results are showing a decrease in the sensitivity by increasing the electrode width. Since the noise level decreases with increasing the electrode area, the ideal design for this flow sensor has a small dimension in the direction of flow (hundreds of nanometers depending on the other design parameters) and a large dimension in the direction perpendicular to it (as large as the channel width).

To investigate the noise level of these sensors, 200 measurements in identical conditions have been done for each data point on the curve. The standard deviations of these 200 identical measurements is defined as the noise level (error bars in the figures). The errors of imaginary and real parts of the impedance seem to be uncorrelated and the real part of the impedance has a noise level of about two times larger than the noise level of the imaginary part (fig. 5). The real part, in addition to the higher noise level, usually has a slightly lower sensitivity than the imaginary part. So, the imaginary part is a better choice for sensing the flow rate. In the results obtained with different device geometries and liquids, we did not observe a meaningful correlation between the flow rate and the noise level, especially in the micro channel device.

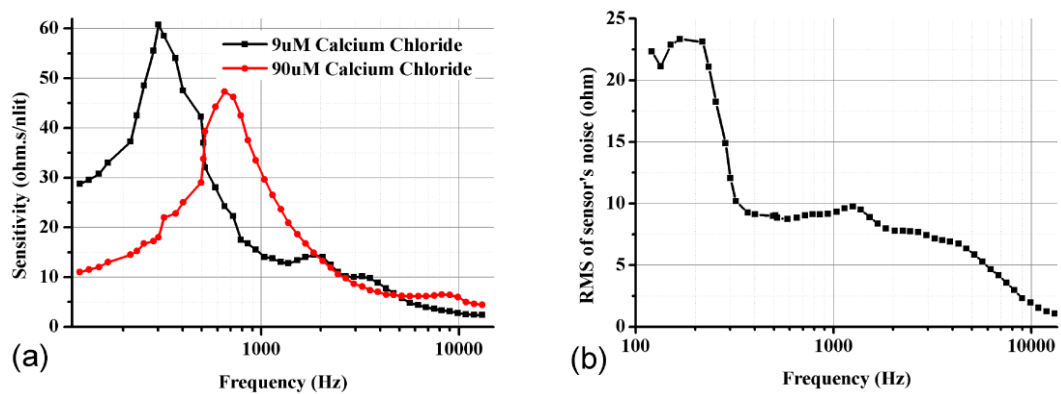

**Figure 7.** a) Sensitivity of a micropore device in different frequencies for different concentrations of calcium chloride. b) The noise level as a function of frequency obtained by 200 measurements at flow rate of 80 nlit/sec at each frequency and calculating their standard deviation. 500 μM calcium chloride in water used in a micropore device with an 80x80 μm$^2$ pore and 50 μm wide electrodes. The applied voltage was a sinusoidal with 50 mV RMS.

As observed in similar experiments (Iliescu et al. 2010), there is a 1/f component in the noise spectrum of the sensor (fig. 7.b) originating from the binding/unbinding of ions to the electrode's surface and capture/emission of electrons at the surface states. The constant level of noise at high frequencies shows the existence of white noise. Since the sensor's signal reduces at high frequencies, the signal to noise ratio is low at very high and very low frequencies. Considering the sensitivity-frequency curve (fig. 7.a) and noise spectrum (fig. 7.b) the best frequency for flow measurement is between hundreds of Hertz to a few kiloHertz depending on the device geometry.

In the micro channel device, after introducing the flow or after changing the flow rate, the impedance response stabilized after less than 60 sec. By waiting for 60 sec before each measurement, we have not observed any meaningful hysteresis (changes in the measured impedance were within the noise level). In micropore devices, the same non-hysteretic behavior was achieved by waiting for about 40 sec.

By increasing the voltage, we have observed an increase in the sensitivity. But, usually it is desired to introduce the minimum chemical reaction by the sensor. So, 50 mV seems to be an appropriate choice of voltage.

## 5. Conclusion

We have demonstrated that an integrated and high sensitivity micro flow sensor can be simply realized in microfluidic devices, by putting two metallic electrodes orthogonal to the flow rate in an insulating channel or pore, and measuring the electrical impedance between the electrodes.

Such a sensor can work on different liquids, while its best performance is in the ionic solutions. These sensors can easily calibrate themselves for an unknown liquid by normalizing the output of the sensor in its linear region. The best impedance parameter to measure the flow rate is the imaginary part of the impedance, due to its lower noise level and the higher sensitivity. The highest sensitivity of these sensors is at mid frequencies ranging from few hundreds of Hertz to few kiloHertz and decreases at very high or very low frequencies. The noise level decreases with the frequency and also decreases with the electrode area. The sensitivity of the sensor increases with the voltage and decreases with the length of the electrode along the direction of the flow. Hereby two novel and easily integratable micro flow sensors introduced, and their sensitivity, noise level, and effects of different design and working parameters on them have been discussed. These easily integratable flow sensors are paving the way for more fully integrated devices with applications in cell culture chips, single cell analysis and many other microfluidic systems. They may also be used to provide a method of flow sensing in many nanofluidics. Such device may finally provide a solution for flow measurement in nano fluidics too.